# Stress tensor mesostructures for freeform shaping of thin substrates

Youwei Yao[1], Brandon Chalifoux[2], Ralf Heilmann[1] & Mark Schattenburg[1]*

[1]Space Nanotechnology Laboratory, MIT Kavli Institute for Astrophysics and Space Research, Massachusetts Institute of Technology; 77 Massachusetts Avenue, Cambridge, MA, 02139, USA.

[2]James C. Wyant College of Optical Sciences, The University of Arizona; 1630 E. University Blvd., Tucson, AZ, 85721-0094, USA.

*Corresponding author. Email: marks@space.mit.edu

## Abstract

Stress-induced shaping, which deforms thin substrates utilizing stressed surface coatings, has enabled and enhanced a host of applications in past decades[1,2,3,4,5,6,7,8,9,10]. Owing to the touchless fabrication process compatible with modern planar technology, the method has been applied from microscale to macroscale applications such as self-assembled micro-structures[5] and space mirrors[11]. However, the deformations created by existing stress-control schemes are limited to certain classes of geometries[12,13] (such as sphere, coma and astigmatism) or rely on boundary constraints and hinges[5] because the stress is unary[12,13], e.g., equibiaxial stress or uniaxial stress with fixed orientation. Here, we present novel stress tensor mesostructures to spatially control the three required stress tensor components, i.e., two normal stresses and a shear stress, over the surface of thin substrates. Three different mesostructure types have been created, each offering distinct advantages. For demonstration, we patterned these mesostructures on the back sides of silicon wafers for freeform shape generation and correction which are not achievable by conventional methods[8,12]. Stress tensor mesostructures will unleash the value of fields related to stress-induced bending from microscale to macroscopy, such as thin freeform substrates that will become increasingly important with the rise of wearable[14,15,16,17] and space optics[18,19,20,21,22].

# Main

Stress-induced bending of thin substrates has recently become an interesting research topic[7,8,9,10,11,12,13,14,15,16,17]. A growing set of techniques to produce controlled deformation and assembly at microscales has resulted in broad impacts on biomedical applications[3], microelectronics[4], micro-electromechanical systems (MEMS)[5], photonics[6], etc. Meanwhile, pertinent techniques at macroscales have led to industry applications and are growing rapidly. Recent development of soft-robotics utilizing stress-based actuation are producing exciting results[8,9]. In addition, research in metasurface optics[14,15,16] which combine meta-lenses with thin freeform mirrors, provides synergy to create ultralightweight folded optical systems. Deterministic bending of thin mirrors is becoming more important as wearable optical systems like augmented reality headsets[17] become ubiquitous, and as more optical systems are launched into space[18,19,20,21,22].

Stress figuring processes create stress on the back surface to bend the substrate and leave the device/optical surface unaffected. Flat or curved substrates can be coated and patterned with high-precision and good throughput using the same technology used to pattern computer chips[9,22]. Stress can be generated using dozens of different coating materials suitable for specific applications, including piezoelectric or magnetostrictive[22,23] films to enable dynamic deformation under external excitations, or metal and oxide thin films[9], to provide stable set-and-forget shaping.

However, the deformations created by existing stress-control schemes are limited to certain classes of geometries (such as sphere, coma and astigmatism) or small clear optical apertures[12,13], because the stress is unary, e.g., equibiaxial stress or uniaxial stress with fixed orientation (Extended Data Fig.1). Generating arbitrary deformation requires fully controlling all three in-plane stress tensor components[12], i.e., the two normal stresses ($\sigma_x$, $\sigma_y$) and shear stress ($\tau_{xy}$). In some cases, boundary constraints and hinges can control deformation of more complicated geometrie[5,24]. Existing methods to provide the required stress control include ion-implantation[25], laser processing[26], and 3D-printing with ferromagnetic or ferroelectric powders[27,28], but these methods rely on difficult-to-control and spatially varying process parameters (ion dose, write speed, droplet size, etc.) that degrade stress manipulation accuracy. Stable methods for precisely controlling the stress state over the back surface of a thin substrates are necessary but still lacking.

Here, we present three methods to fully control the stress state over the entire substrate back surface using carefully engineered mesostructures that rely primarily on patterning accuracy. We designed and patterned three types of mesostructures on the backside of 100 mm silicon wafers to demonstrate high-precision, free-standing shape generation and correction. The demonstrated methods can be directly applied to manufacture thin optical mirrors similar to the way computer chips are patterned. In addition, the patterns can easily be scaled down for microscales, and the stress provider can be switched to functional materials for active deformation.

## Generation of uniaxial stress

A state of plane stress (two normal stresses and a shear stress) can be decomposed into equibiaxial stress ($\sigma_{equi}$) and a uniaxial stress ($\sigma_{uni,j}$) component at a set of specific orientations ($\phi_j$) by inverting Eq. (1) below.

$$\begin{bmatrix} \sigma_x \\ \sigma_y \\ \tau_{xy} \end{bmatrix} = \sum_{j=1}^{m} \begin{bmatrix} \cos^2(\phi_j) & \sin^2(\phi_j) & 2\cos(\phi_j)\sin(\phi_j) \\ \sin^2(\phi_j) & \cos^2(\phi_j) & -2\cos(\phi_j)\sin(\phi_j) \\ -\cos(\phi_j)\sin(\phi_j) & \cos(\phi_j)\sin(\phi_j) & \cos^2(\phi_j)-\sin^2(\phi_j) \end{bmatrix} \begin{bmatrix} \sigma_{equi} + \sigma_{uni,j} \\ \sigma_{equi} \\ 0 \end{bmatrix} \quad (1)$$

The term on the left is a stress state represented in x-y coordinates. The terms on the right are the product of a 3 by 3 rotation matrix and a stress state vector in the principal coordinate frame (Extended Data Fig. 1). The summation limit $m$ depends on the mesostructure type and will be explained later.

Equibiaxial stress already exists in conventional coatings. Here we present a novel scheme to create uniaxial stress as follows and illustrated in Figs. 1**a**. First, we coat a substrate with a film of equibiaxial stress. Second, we pattern the coated surface with grating lines in which trenches are extended into the substrate, in contrast with previous research that patterned the film only[29]. Our study has revealed that uniaxial stress can be generated with a grating pitch close to the total trench depth, which can be significantly greater than the thickness of the film. This relaxation of the requirement on the grating pitch allows the patterns to be more easily manufactured.

Fig. 1**a** illustrates an example of grating lines patterned horizontally (x direction) on the backside of a 100 mm-diameter, 0.5 mm-thick silicon wafer. The pitch is 10 µm, much larger than the thickness of the stress provider—a 200 nm-thick thermally-grown silicon dioxide (TOx) layer with -70 N/m equibiaxial compressive stress. By using a 2D finite-element (FE) model (Extended Data Fig. 2), we determined how the local bending curvature of the wafer surface in the y direction varies with the trench depth in silicon. After modelling, it is instructive to assume the grating structure can be replaced by a fictitious film of uniform stress, wherein the equivalent uniaxial stress of the film in the y direction is calculated from the wafer curvature for different aspect ratios (AR=h/w, and thickness of the TOx is negligible) of the grating teeth, and then normalized to one for an AR of zero. The results are plotted as the red solid line in Fig. 1**b**, suggesting the equivalent film stress in the y direction drops to zero when the AR is around 0.3. In addition, a surprising stress reversal occurs when the AR is higher than 0.3, reaching a maximum when AR=0.5.

To test the modeled results, we patterned uniform grating lines with deep trenches on the backsides of five silicon wafers. Each wafer had the same 10 µm grating pitch, but the trench depths produced by deep reactive ion etching varied with ARs from 0 to 1. The wafer deformation induced by the grating patterns was monitored by a Shack-Hartmann (S-H) metrology tool. Based on these measurements (Extended Data Fig. 3), the normalized equivalent stresses in the y direction were calculated, with results plotted in Fig. 1**b** (red squares), matching the modeled results (solid red line). We have concluded that controlled uniaxial stress can be created when the AR is around 0.3. In addition, AR>1.0 can also produce a nearly uniaxial stress.

We also investigated the counterintuitive stress reversal (from compressive to tensile) when the AR is between 0.3 and 1.0. We measured the ratio of the curvatures between x and y directions on the fabricated wafers (blue circles), comparing the results with a classic model (blue line) which has a similar configuration but assumes the trenches are only in the film[29]. The deviation suggests to avoid the AR between this region. More interestingly, a counterintuitive negative bending could occur without changing the coating stress. To verify this result, patterns of cylindrical pillars with AR=0.5 and AR=0.25 were trenched on the backsides of two silicon wafers (Fig. 1**c, e**). Equibiaxial stresses are achieved due to the symmetric structures. The measured deformations confirm the counterintuitive negative bending (Fig. 1**d**, **f**), which adds the possibility of bimorph deformations for more applications.

## Control of stress tensors with mesostructures

Based on this new capability of generating controlled uniaxial stress, we have created three types of periodic mesostructures arrayed in a 2D lattice on the backside of silicon wafers to produce stress tensors for freeform surface shaping. This scheme assumes that the mesostructures are comparable to, or smaller than the substrate thickness, so that each unit cell can essentially be considered a pixel of controllable stress representing a continuous tensor stress field. Fig. 2**a** shows the Type-I mesostructure arrayed in a hexagonal lattice of 500 μm horizontal spacing. The highlighted yellow disk of diameter A represents a pattern of the TOx layer. The ratio of disk area to unit cell area is controlled to determine the local magnitude of equibiaxial stress. The parallel lines with 10 μm pitch represent trenches with AR=1 through the TOx into the silicon, which convert the stress within the grating area from equibiaxial to uniaxial, as discussed previously. The width B of the grating region controls the local magnitude of uniaxial stress, while the spin angle $\phi$ controls its local orientation (see Eq. 1). Figs. 2**b-d** depict the surface deformations effected by the different pattern components.

After determining the target surface deformation, the desired stress is transformed from global orientation ($\sigma_x$, $\sigma_y$, $\tau_{xy}$) to local orientation ($\sigma_{equi}$, $\sigma_{uni,j}$, $\phi_j$) using Eq. 1. For Type-I structures, m=1 and $\phi_1$ is the orientation of the grating lines, $\sigma_{equi}$ is the equibiaxial stress and $\sigma_{uni,1}$ is the uniaxial stress. These local stress states can be transformed, in turn, into 2D distributions of local geometric parameters A, B and $\phi$ using a calibration process (Extended Data Fig. 4). Tensile and compressive stress components can be achieved by adding a bias film on the front and/or back surfaces (Extended Data Fig. 5).

The Type-I structures can produce deformations efficiently since grating lines in each unit cell are rotationally aligned to maximize the principal stresses. However, secondary adjustment of the stress orientation for post-correction deformation or shape actuation would be difficult. For this reason, we introduced the Type-II mesostructure as shown in Fig. 2**e**. The unit cell is highlighted and indicated by magenta dashed lines consisting of a triplet of unique patterned circles. In this pattern, m=3 in Eq. 1 and $\phi_j$ is one of the three grating line orientations (60°, 0° or -60°), $\sigma_{equi}$=0, and $\sigma_{uni,j}$ is the uniaxial stress (recall that j indicates the location of the cell on the substrate). In contrast with the Type-I mesostructure, the geometric variables of Type-II are the diameters of the TOx disks at the three orientations (C, D and E in Fig. 2**e**), which can manipulate the stress tensor components for orthogonal deformations (Figs. 2**f-i**) without changing the grating line orientations. This configuration can enable secondary adjustment or active control of the stress tensor if the stress provider is a ferromagnetic or piezoelectric material. Although Type-II has more flexibility, the magnitude of the stress generated is less than Type-I since only one third of the fractional TOx area is in effect at an arbitrary orientation.

We also developed a Type-III structure depicted in Fig. 2**j** for the purpose of producing higher stresses with comparable flexibility. The Type-III structure has grating lines with TOx also coated on the side walls and floors of grating teeth (Fig. 2**k**), providing additional uniaxial stresses proportional to the AR of the grating structure. Sidewall coatings can be realized by isotropic deposition methods after etching of grating lines, such as thermal oxidation or atomic layer deposition (ALD). However, these methods coat the area between grating teeth at the same time, producing undesired equbiaxial stress that can counteract the desired deformation.

Therefore, the configuration of the Type-III structure was modified to resolve the problem. The diameters of the grating disks are controlled as variables for desired uniaxial stresses. The TOx disk patterned after deposition is slightly larger than the grating circles. The marginal ring of the TOx in each disk, indicated by F, produces an additional equibiaxial stress to even out the variation of equibiaxial stress generated by grating disks. Therefore, the equibiaxial stress within each disk is constant, which can be compensated by a uniform coating from the other side of the thin substrate for the initial demonstration in this work.

**Results**

To demonstrate the effective use of a Type-I structure for figure correction, we patterned the backsides of two free-standing silicon wafers to generate deterministic freeform deformations. We selected the Zernike trefoil as a deformation target since it requires uniaxial stress with varying magnitude and orientation (See Methods), which requires the manipulation of stress tensors beyond the conventional equibiaxial type.

Fig. 3**a** shows a portion of the structure design for trefoil deformation near the substrate center. A 200 nm-thick TOx layer was used as a stress provider. Fig. 3**b** shows 25 microscope images of individual unit cells evenly distributed within a 70 mm by 70 mm area. The measured deformation is represented by 12 Zernike polynomial terms of which the coefficients are plotted by the green triangles in Fig. 3**g**. Fig. 3**c** shows the measured deformation, which demonstrates that trefoil has been successfully generated.

Super flat thin substrates are in demand for wide applications such as semiconductor manufacturing and metamaterial lens. For a final demonstration, we selected a silicon wafer and then patterned an engineered Type-I backside structure designed to flatten the surface. Previous studies have shown that generating deformation with different Zernike terms requires different types of stresses. Among the 12 terms shown in Fig. 3**g**, six of them need antibiaxial stress, four need equibiaxial stress and two need a combination of both (Extended Data Fig. 1). In this work, flattening surfaces by simultaneously minimizing 12 terms is a powerful demonstration of stress tensor manipulation. The measured surface before correction is the S shape in Fig. 3**f**, with ~10 μm P-V, corresponding to the coefficients plotted by the magenta squares in Fig. 3**g**. After calculation of the required stress tensor distribution based on an analytical solution and an FE model and taking into account the anisotropic elastic properties of silicon[30], Type-I mesostructures are designed and patterned on the backside to provide adequate flattening stress tensors. Fig. 3**e** shows the area between two unit cells, where the trenched lines are clearly visible. The flattened profile in Fig. 3**f** shows the measured surface after patterning, corresponding to the blue circles in Fig. 3**g**, representing a RMS slope improvement factor of 27 (RMS height improvement factor of 21).

These results can benefit applications which only need one-time shape forming. Fig. 4 shows results conducted by Type II and III structures which allow secondary forming and possibly active actuation. Fig. 4**a** shows a microscope image of a patterned Type-II structure for trefoil deformation. The measured deformation in Fig. 4**b** shows the result is not perfect. As indicated by the arrow, a side-lobe in the x direction is slightly higher, which is caused by an undesired astigmatism generated by the patterning process. This astigmatism component is clearly shown in the magenta circles in Fig. 4**g**. Since the generated stress tensors can be adjusted after patterning due to the flexibility of Type-II structures, we performed a secondary exposure to remove the undesired astigmatism. The TOx at the center of the disks, where the grating lines are along the x direction, was removed iteratively (See Methods), but not etching into the silicon (Fig. 4**c**). Fig. 4**d** shows the measured deformation after the secondary patterning, where the side-lobe amplitude has been suppressed. The Zernike coefficients plotted by the blue circles in Fig. 4**g** confirm that the generated trefoil has improved. The capability of secondary adjustment granted by the Type-II structure can benefit the precision of the freeform process. In addition, the structure can enable actively controlled stress tensors for more applications, by switching the stress provider from TOx to ferromagnetic or piezoelectric materials.

Despite these advantages, comparing Figs. 3**c** and 4**d**, the deformation amplitude produced by Type-II structures is only one-third of that produced by Type-I due to the inefficient nature of the design. Fig. 4**e** shows a Type-III structure for higher deformation amplitude with the same flexibility as Type-II. Since the AR of the grating lines is 1.0, which enables three times the coating area, the generated deformation in Fig. 4**f** shows an amplitude ~3 times higher than Type-II, as expected.

## Summary

We have presented three types of novel mesostructures that can manipulate the general stress tensor on thin substrates to create high-precision freeform deformations not achievable by conventional methods. The mesostructures are compatible with modern planar fabrication technologies, which have the potential to be scaled for a wide range of applications. In addition, the Type II and III mesostructures allow for secondary forming, which could extend the capability of deformation from static to dynamic in the future.

**Fig. 1: Unusual bending effect generated by high aspect ratio surface structures.**

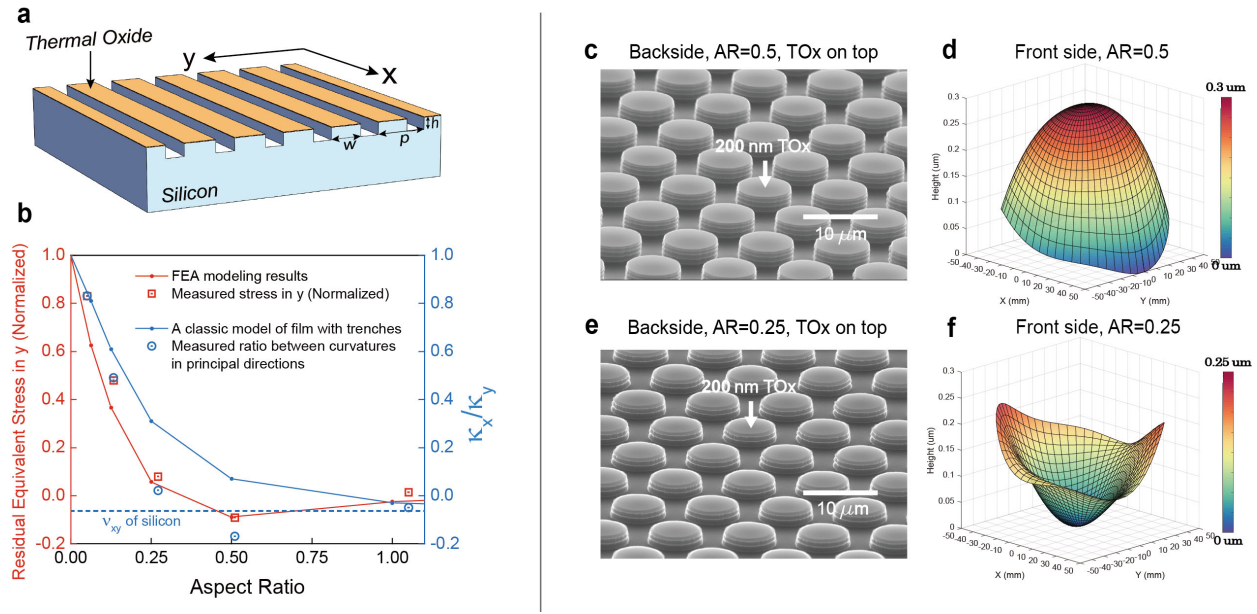

**a**, Grating lines trenched into silicon substrates, with the TOx coated on the top of the grating tooth. AR=h/w, thickness of TOx is negligible. **b**, The red line and squares represent the modeled and measured equivalent stress (normalized by the stress when AR=0) in the y direction, assuming the grating structure in (**a**) is a fictitious equivalent continuous film. The blue line and circles show the ratio of the curvatures between x and y directions on a grating-patterned silicon wafer. The blue line shows a theoretical result from a classic model, in which the trenches are only in the film[29] (AR=h/w for teeth of coatings, width comparable with the coating thickness). The blue circles are the measured results (AR=h/w for teeth of trenched silicon, width much larger than the coating thickness). The dashed blue line is the asymptotic line of the solid line derived from the classic model, with the value of silicon's Poisson's ratio, $\nu_{xy}$. **c**, SEM image of a cylindrical pillar structure with AR=0.5, patterned on the backside of a silicon wafer. 200 nm of TOx (compressive stress) is coated on the top of the pillars. **d**, Measured deformation on the front side of the silicon wafer, demonstrating an equivalent tensile stress on the backside when AR=0.5. **e**, **f**, Results for the case where AR=0.25.

**Fig. 2: Sketches of three types of mesostructures laid out in a hexagonal lattice and the deformations of silicon wafers generated by different structural components.**

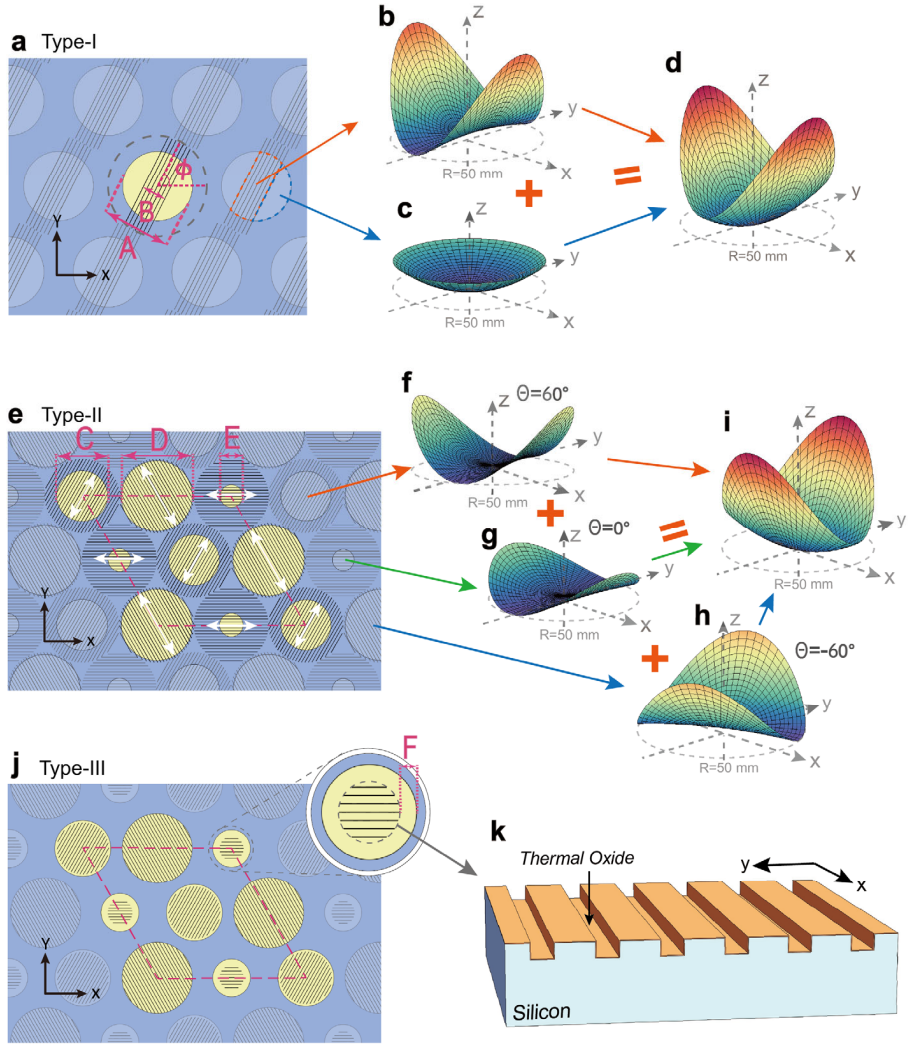

**a**, Type-I mesostructure with a highlighted exemplar unit cell at the center. The yellow disk is a 200 nm-thick thermal oxide layer. The parallel lines are trenched gratings through TOx into the silicon substrate. The grating line pitch is 10 μm and the aspect ratio of the grating tooth is 1. Parameters A, B and $\phi$ define the dimensions which can vary from cell-to-cell within the boundary which is 500 μm-diameter circle shown by the gray dashed line. **b**, **c**, Sketch of deformations generated by the different parts of the Type-I structure, assuming all of the unit cells have the same geometry. The grating area creates a taco-shell profile (**b**) while the remaining unpatterned oxide area creates a spherical shape (**c**). **d**, The sum of these two deformations, which is a combination of sphere and astigmatism represented by Zernike polynomials (**b** + **c** = **d**), creates a new structure. **e**, Type-II mesostructure with highlighted unit cell indicated by magenta dashed lines. The diameter of each grating circle area is 500 μm and the grating line pitch is 10 μm. The orientations of the grating lines are fixed at -60°, 0°, 60°, as noted by the white arrows. Parameters C, D and E represent TOx disk diameters that have different grating orientations. **e-h** Sketches of the deformations generated by the grating circles, assuming all of the unit cells have the same geometry. **i**, Combining deformations to create a target deformation (**f** + **g** + **h** = **i**). **j**, Type-III mesostructures are similar to Type-II except the TOx disks are larger than the circled grating areas to provide additional equibiaxial stress by adjusting parameter F. **k**, The grating lines are trenched before the TOx coating and patterning process. Note the side wall and the adjacent area of the grating teeth are also coated with TOx. As a result, the magnitude of the uniaxial stress produced by the grating area is increased, proportional to the AR.

**Fig. 3: Freeform shape generation and correction of silicon wafers conducted by Type-I mesostructures patterned on back surfaces.**

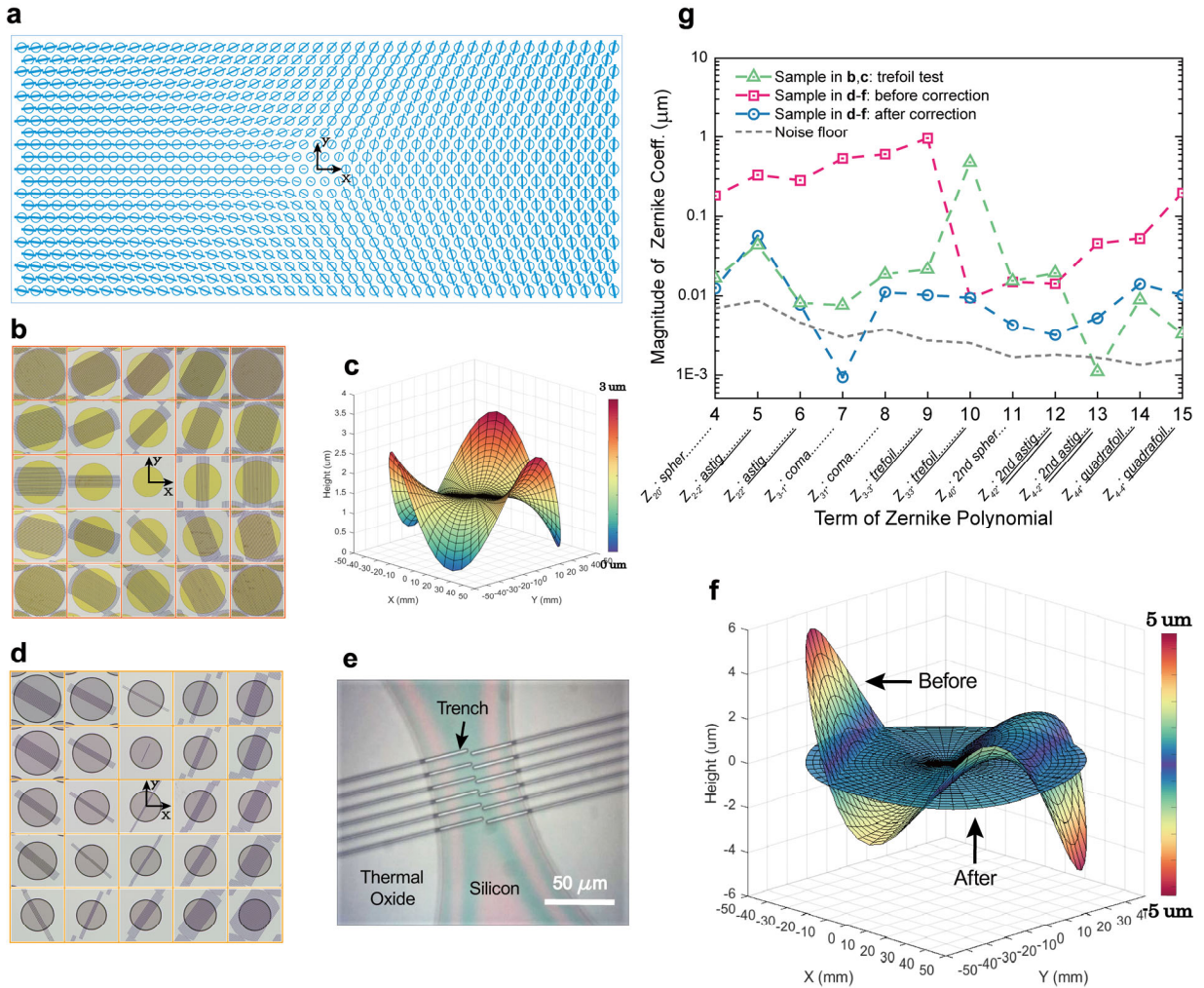

**a**, A portion of the design for generating trefoil deformation. The pitch between two adjacent circles is 500 μm. **b**, Microscope images of 25 fabricated individual unit cells for trefoil deformation. The cells are evenly distributed within a 70 mm by 70 mm area. **c**, Measured wafer trefoil deformation. **d**, Microscope images of 25 unit cells for flattening a wafer's surface. Note: The TOx disks are darker than those in (**b**) due to a difference of thickness. **e**, Microscope image of the area between two unit cells. **f**, Measured wafer surface before (S shape) and after (flat shape) the patterning process. **g**, Zernike coefficients of the measured deformations and surface profiles in (**c**) and (**f**). The dashed gray line represents the repeatability of our S-H metrology tool. The tilted text on the bottom indicates the IDs of the 12 Zernike terms. Letters without underline indicate terms that can be generated by equibiaxial stress, single underline for terms that can be generated by antibiaxial stress, and double underline for terms that require a combination of equibiaxial and antibiaxial stresses.

**Fig. 4: Trefoil deformation of 100 mm silicon wafers generated by Type-II and Type-III mesostructures patterned on back surfaces.**

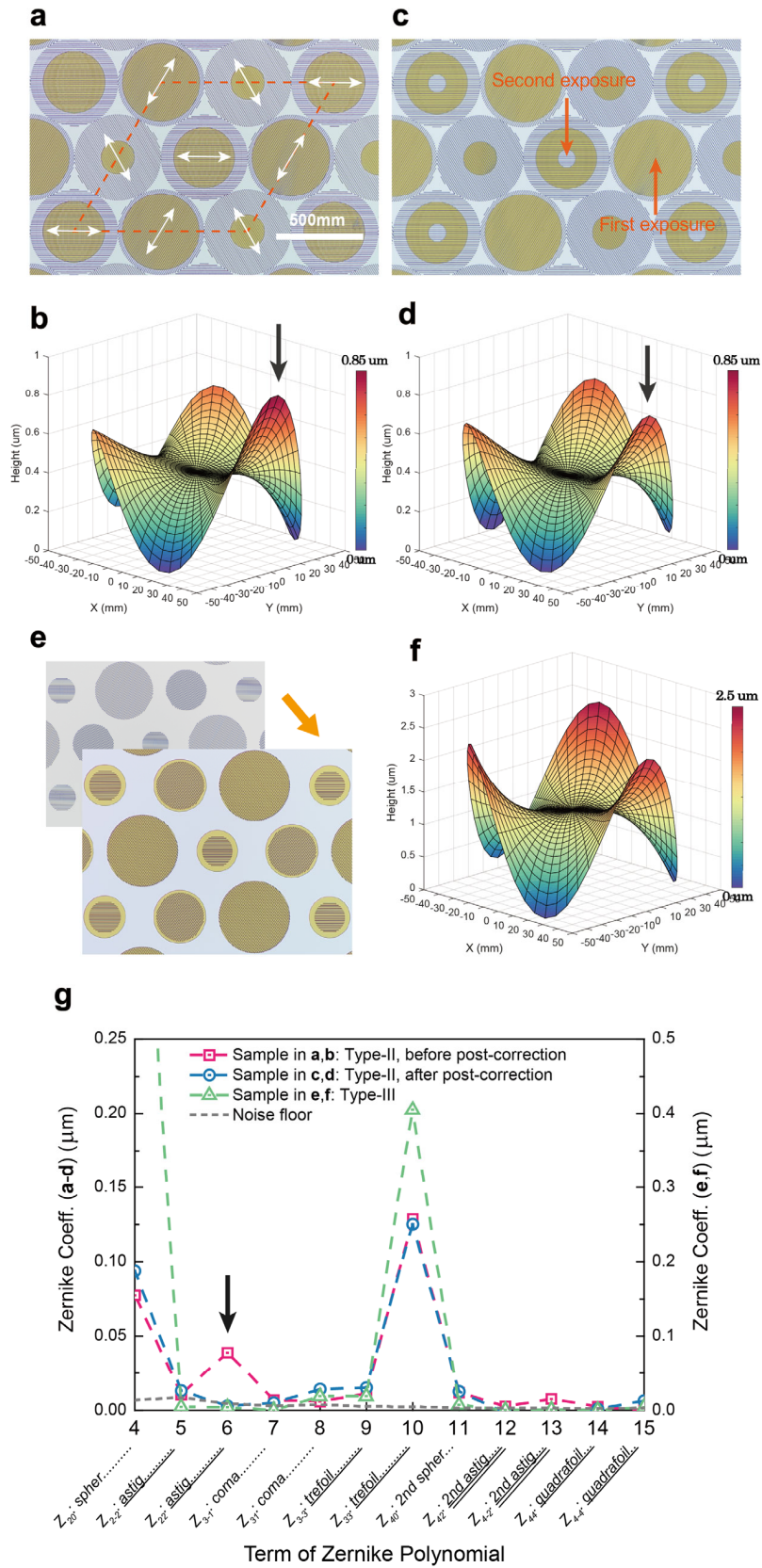

**a**, Microscope image of patterned Type-II mesostructures for trefoil deformation. Dashed lines indicate a unit cell and the white arrows note the orientation of grating lines. **b**, Measured trefoil deformation generated by (**a**). The dark arrow indicates one of the generated side lobes is higher than expected, created by an undesired astigmatism term. **c**, Microscope image of a Type-II mesostructure after secondary exposure. A portion of the TOx in grating disk centers with specific orientations are removed. **d**, Measured trefoil deformation after the secondary exposure, showing the side lobe indicated by the dark arrow is suppressed. **e**, Microscope images of Type-III mesostructures. The background image is the silicon surface trench with grating lines before coating and patterning. The picture in the foreground shows patterned TOx layers on the top and the side of the grating tooth. **f**, Measured trefoil deformation created by (**e**). The amplitude of the deformation is ~3X higher than the one produced by Type-II. **g**, Measured deformations of 12 Zernike coefficients. The dashed gray line represents the repeatability of our S-H metrology tool. The tilted text indicates the IDs of Zernike terms, in the same manner as Fig. 3**g**. The dark arrow indicates an undesired astigmatism term that showed up in the first exposure of Type-II structure.

# Methods

## Limitation of deformations created by equibiaxial stress fields

Based on a previous study[12], conventional coating stress on flat wafer substrates can only exactly generate a restricted set of Zernike deformations (denoted as $Z_{nm}$ where n is the radial degree and m is the azimuthal order) over the full aperture of a substrate, such as sphere (defocus) ($Z_{20}$), primary coma ($Z_{31}$ and $Z_{3-1}$) and 2nd order sphere ($Z_{40}$), due to the fact that stress fields in conventional coatings are equibiaxial. Extended Data Fig. 1a illustrates the Zernike polynomial deformations and the corresponding stress fields required to achieve them. For the components in red and yellow and using equibiaxial stress only, approximate correction over a full aperture or exact correction over a sub-aperture are possible, but exact correction over a full aperture is not. For example, the generation of the astigmatism term ($Z_{22}$) needs an antibiaxial stress with fixed magnitude and orientation (Eq. 2). In addition, the generation of trefoil ($Z_{33}$) requires antibiaxial stress, but the magnitude and orientation vary with location as demonstrated in Eq. 3.

$$N_1 = S_0, \quad N_2 = -S_0, \quad \phi = 0 \tag{2}$$

$$N_1 = S_0 \times \frac{r}{R}, \quad N_2 = -S_0 \times \frac{r}{R}, \quad \phi = \frac{1}{2}(\pi - \theta) \tag{3}$$

Here $N_1$ and $N_2$ are the local principal stresses with orientation of coordinate $\phi$, $S_0$ is a constant, and R is the radius of the substrate. The stresses are expressed in polar coordinates where $(r, \theta)$ represents the location on the substrate. Extended Data Fig. 1**b** shows a sketch of the local stresses in these coordinates.

## 2D FE modeling of deformations created by trenched lines on coated surfaces

We developed 2D FE models in Abaqus to analyze the deformation created by trenched lines on coated surfaces. Extended Data Fig. 2**a** and 2**b** depict the local geometry, deformations and stress distributions of the model with AR=0.5. FE meshes are visible on the plot. Since the models are two-dimensional, the element type is set as plane strain for analyzing the cross sections. Extended Data Fig. 2**c** shows calculated deformations for different ARs. The curvature of each profile represents the equivalent stress in the fictitious films along the y direction (direction normal to the grating lines). The normalized stresses are calculated by using the wafer with AR=0 as reference. Results are plotted with a solid red line in Fig. 1**b**.

## Metrology and repeatability tests

In this work we use a Shack-Hartmann metrology tool[31] to measure the deformation of 100-mm-diameter, 0.525 mm-thick silicon wafers generated by stress patterns. The tool is enclosed in a wind shield structure to mitigate the influence of turbulence. The wafer mount is carefully designed to hold the wafer vertically for minimum distortion[32]. A nickel-coated aluminum block with surface flatness of $\lambda/4$ is integrated into the wafer mount to serve as a reference surface. The measured raw data is fitted by 36 terms of Zernike polynomials. A silicon wafer was mounted, measured, and removed from the tool 10 times to test repeatability which is below 10 nm in RMS of each Zernike term. In the test, the mean coefficient of each Zernike term is used as a reference. The first three terms are omitted since they represent offset, tip and tilt, which have no impact on the measured profile.

**Five wafers patterned by gratings with different aspect ratios (Fig. 1b)**

To demonstrate the effect of grating lines trenched into silicon, we have patterned five silicon wafers (diameter 100 mm, thickness 0.525 mm, crystal orientation <100>) with fixed pitch (10 µm) and different aspect ratios (ARs from 0 to 1). The fabrication process is described in Supplementary information.

Based on the measured deformations, we used 3D finite element (FE) models to fit the profiles assuming the deformations are created by fictitious stressed films on the wafer backsides. Details of the 3D FE model fitting process are described in previous work[33]. For the wafer with AR=0 (Extended Data Fig. 3**a**), the stress is equibiaxial. The curvature of deformations before and after resist patterning/TOx etch are measured to calculate the area fraction of the remaining TOx which is determined by the width of the trenches in TOx layers. The area fraction is used to derive the equibiaxial stresses of each wafer (Step 4) before DRIE. For the wafers with AR from 0.05 to 1.05 (Extended Data Fig. 3**b-f**), the deformations were fitted by astigmatism ($Z_{22}$) and sphere ($Z_{20}$) terms.

The principal stresses along the y direction were calculated based on the FE models and then normalized by the derived equibiaxial stresses before DRIE. The results are depicted by the red squares in Fig. 1**b**. The ratio of the curvatures between x and y directions are also calculated and plotted in Fig. 1**b** (blue circles). Note: a negative curvature along the y direction appears in Extended Data Fig. 3**e**, which is a direct illustration that tensile stress is imparted when AR=0.5.

**Cylindrical pillars patterned on the backsides of silicon wafers**

Grating trenches in silicon with AR=0.5 create a counterintuitive stress reversal as demonstrated by Fig. 1**b** and Extended Data Fig. 3**e**. However, a compressive uniaxial stress can generate a negative curvature along the perpendicular direction due to the Poisson's ratio of the substrate[30], which is indistinguishable from the effect of a tensile stress. The negative curvature in the y direction for AR=0.5 (Extended Data Fig. 3**e**) is higher than that for AR=1 (Extended Data Fig. 3**f**), which suggests an abnormal bending moment. In order to validate this prediction, we patterned cylindrical pillars trenched into silicon. Since the structure of cylindrical pillars is axially symmetric, the stress reversal due to the structure of AR=0.5 would appear equibiaxially. The equibiaxial stress-induced deformation (sphere) can be clearly observed and compared to the results derived from Stoney's equation. Figs. 1**c** and 1**d** clearly show that a toothed structure with a stressed top coating can generate stress reversal when AR=0.5. Although the cylindrical pillars are not in the same configuration as the one-dimensional grating trenches, these results directly demonstrate stress reversal induced by the tooth-structure when AR=0.5.

In addition, the solid blue line in Fig. 1**b** is a result of a classical model which assumes the trenches are only in the film. The AR in the classical model is defined as the height of the coating tooth versus the width. The classic model is derived from a two-dimensional Stoney's equation and assumes the trench gaps are comparable to the coating thickness[29]. The deviation between the blue circles and line in Fig. 1**b** demonstrates that the classic model cannot explain the stress reversal results.

**Calibration of Type-I mesostructures**

The Type-I mesostructure generates equibiaxial and uniaxial stresses simultaneously within each unit cell. Therefore, the trenched lines can lead to a relaxation of generated equibiaxial and uniaxial stresses, as depicted by Extended Data Fig. 4, which can compromise stress control precision. Although Type-I provides the highest magnitude of stress among the three types, it requires a calibration process to connect the geometric parameters and the generated stresses. We define two parameters, 1) Duty Cycle (DC) and 2) Area Fraction (AF) (Extended Data Fig. 4**a**), which can be mapped to parameters A and B shown in Fig. 2**a**. The DC represents the ratio of the area between the TOx disk and the unit circle (gray dashed line in Extended Data Fig. 4**a**). The AF is the area ratio between the grating-trenched area and the TOx disk. As a result, by controlling these two parameters, the magnitude of the uniaxial stress and equibiaxial stress can be manipulated. In order to achieve good precision, we performed the calibration process illustrated in Extended Data Fig. 4.

We fabricated 20 wafers for testing different combinations of two-dimensional geometric parameters. Type-I mesostructured patterns with fixed DC, AF and grating rotation angle (horizontal) were uniformly patterned on the backside of each individual wafer. The DC and AF vary for different wafers from 0% to 100% for testing deformation creation. The topologies of the 20 wafers were monitored for comparison with calculated equibiaxial and uniaxial stresses based on the 3D FE model[33]. Extended Data Fig. 4**b** shows microscope images of the unit cell regions from 16 wafers with different DCs and AFs. Extended Data Fig. 4**c** depicts the corresponding measured deformations of those wafers. (The four samples with AF=0 were omitted since they only have equibiaxial stress and spherical deformations.) The 20 measured deformations were fitted by 3D FE models. The generated equibiaxial and uniaxial stresses are normalized by the value of equibiaxial stress when DC = 100% and AF = 0%. Results are plotted in Extended Data Figs. 4**e** and 4**f** which are used as calibration maps. By use of calibration map look-up tables, the model can determine the required DC and AF values to achieve target equibiaxial and uniaxial stresses and thus achieve desired deformations.

The fabrication process for the 20 wafers is described in supplementary informations.

**A proposed solution for bimorph deformation with two kinds of stressed coating materials**

A possible solution for bimorph deformation is to use two different stress providers on one substrate. For Type-I, the material of the stressed coating with each grating disk needs to be selected carefully. Extended Data Fig. 5 shows a possible arrangement of two stressed coatings for the Type-II mesostructure. A candidate material for providing tensile stress could be a ceramic such as silicon nitride[34]. If the magnitude of the stress in each material is the same, the spatially averaged stress of the patterns depicted in Extended Data Fig. 5 should be zero. We expect mesostructures such as the one in Extended Data Fig. 5 could be integrated on the backside of new silicon wafers in future, thus bi-directional topology control could be achieved after the front side is patterned with device and coating films.

**Generation of trefoil by Type-I mesostructured (Figs. 2a-2c)**

In this demonstration, the backside stress provider is a TOx layer which intrinsically provides ~-300 MPa compressive stress. Since the generation of trefoil needs tensile stresses as demonstrated in Eq. 3, we decided to use a uniform TOx layer on the frontside to bias the shape. We assume the uniform compressive stress in the front side TOx is $S_{front}$, therefore the same compressive stress on the back side needs to be added in Eq. 3 to counteract the bias of the shape while generating the trefoil. When $S_{front}$ is equal to $-S_0$, the stress field on the backside for generating the trefoil deformation is compressive everywhere, as demonstrated by Eq. 4.

$$N_1 = -S_0 \times (1 - \frac{r}{R}), \quad N_2 = -S_0 \times (1 + \frac{r}{R}), \quad \phi = \frac{1}{2}(\pi - \theta) \qquad (4)$$

The combination of two principal stresses in Eq. 4 can be translated to a combination of equibiaxial and uniaxial stresses, expressed as follows.

$$S_{equibiaxial} = -S_0 \times (1 - \frac{r}{R}), \quad S_{uniaxial} = -2S_0 \times \frac{r}{R}, \quad \phi = \frac{1}{2}(\pi - \theta) \qquad (5)$$

$S_0$ represents the maximum equibiaxial stress (when r=0 which is at the center of the wafer), and $2S_0$ stands for the maximum uniaxial stress (when r=R which is at the rim of the wafer).

The calibration maps (Extended Data Fig. 7) show that the maximum uniaxial stress is 66% of the magnitude of the equibiaxial stress when AF=1 and DC=1. Therefore, Eq. 5 is modified as,

$$S_{equibiaxial} = -0.33 N_0 \times (1 - \frac{r}{R}), \quad S_{uniaxial} = -0.66 N_0 \times \frac{r}{R}, \quad \phi = \frac{1}{2}(\pi - \theta) \qquad (6)$$

where $N_0$ is the equibiaxial stress when AF=0 and DC=1. Based on Eq. 6 and the calibration maps, the distributions of the DC, AF and the spin angle for the trefoil deformation can be determined to generate patterns for fabrication (Fig. 3**a**).

The fabrication process is very similar with the one used for the calibration process, the only difference being the presence of a TOx layer on the front side. We preserve the front side TOx all through the steps until the last one – when the mesostructures are successfully patterned on the backside, and we perform an iteration process to uniformly reduce the thickness of the TOx layer on the front side until the measured spherical term ($Z_{20}$) reaches a minimum. Extended Data Fig. 6 shows the evolution of the deformation when the TOx thickness on the front side is reduced. The iteration process is described in supplementary information.

**Shape correction with the Type-I mesostructure (Figs. 3d-3f)**

The tensor fields for producing deformations of the first 15 Zernike terms individually are listed in Extended Data Table 1, which are derived from an analytical solution[12]. In order to flatten silicon wafers with initially curved profiles as shown in Fig. 3**f** (S-shape), the stress field, which is a linear combination of the tensors in Extended Data Table 1, should be determined and patterned on the wafer backsides in accordance with the Zernike coefficients of the measured deformation. However, for these tests we used monocrystalline silicon wafers with <100> orientation where the Young's modulus of the material is orthotropic[30]. Therefore, the analytical solution based on substrates with isotropic material properties needs modification. In order to study the influence of the anisotropic properties, we established a FE model with the setting of orthotropic material. The model is a silicon wafer with a coating on the backside. The stress tensor field can be assigned in the coating for calculating the deformation. By using the stress tensors in Extended Data Table 1, we attempt to generate the deformation of each Zernike term with the coefficient of 500 nm in the model. Results are shown in Extended Data Table 2.

The modeling results show that some of the Zernike terms are influenced by the orthotropic substrate properties significantly. For example, for the Zernike term #13, which is 2nd order astigmatism $Z_{4-2}$, the generation of a 500 nm coefficient leads to 575 nm deformation of $Z_{4-2}$ and 889 nm deformation of $Z_{2-2}$ (astigmatism). Therefore, the correction based on the stress fields for isotropic substrates can result in low precision.

Our solution for this problem is expressed in Eq. 7. The FE modeling results are used as a matrix to compensate the target deformation.

$$\frac{1}{508} \cdot \begin{pmatrix} 508 & 0 & 0 & 6 & 1 & -1 & 0 & -248 & 13 & 0 & 155 & 0 \\ 0 & 791 & 0 & -2 & -14 & 0 & -3 & 0 & 0 & 889 & 0 & 6 \\ 0 & 0 & 509 & 17 & -3 & -3 & -1 & -44 & 57 & 0 & -5 & 0 \\ 0 & 0 & 0 & 543 & 0 & 36 & 0 & 8 & -6 & -1 & 1 & 0 \\ 0 & 0 & 0 & 0 & 548 & 0 & -35 & 1 & 1 & -6 & 0 & -2 \\ 0 & 0 & 0 & 108 & 0 & 629 & 0 & 36 & -30 & -2 & 4 & 1 \\ 0 & 0 & 0 & 0 & -122 & 0 & 626 & -7 & -5 & 12 & -1 & 3 \\ 0 & 0 & 0 & 1 & 0 & 0 & 0 & 564 & 1 & 0 & -41 & 0 \\ 0 & 0 & 0 & -1 & 0 & 0 & 0 & 4 & 506 & 0 & 0 & 0 \\ 0 & 0 & 0 & 0 & -2 & 0 & 0 & 0 & 0 & 575 & 0 & 1 \\ 0 & 0 & 0 & -7 & -1 & 1 & 0 & -148 & -15 & 0 & 635 & 0 \\ 0 & 0 & 0 & -1 & 4 & 0 & 1 & 0 & 0 & -6 & 0 & 618 \end{pmatrix} \begin{pmatrix} C_4 \\ C_5 \\ C_6 \\ C_7 \\ C_8 \\ C_9 \\ C_{10} \\ C_{11} \\ C_{12} \\ C_{13} \\ C_{14} \\ C_{15} \end{pmatrix} = \begin{pmatrix} 184.06 \\ 334.09 \\ 287.18 \\ -537.99 \\ -606.09 \\ -958.03 \\ 9.39 \\ -14.84 \\ 14.13 \\ -45.85 \\ 52.6 \\ -199.14 \end{pmatrix} \quad (7)$$

Here $C_4 - C_{15}$ are the compensated Zernike coefficients derived from the measured deformation (S-shape in Fig. 3**f**) which is represented by the coefficients on the right side of the equation. By using compensated Zernike coefficients and the library of the stress tensors in Extended Data Table 1, the tensor fields for flattening the profile of the silicon wafer are calculated and depicted in the Extended Data Fig. 7.

Based on this calculation, the required initial stress of the TOx layer on the backside before patterning was -464.8 N/m, corresponding to a thickness of 1.414 μm. The stress on the front side is -126.3 N/m which requires a 0.383 μm-thick TOx layer.

The calculated stress fields were converted to the maps of DC and AF by looking up the calibration maps in Extended Data Figs. 4d and 4e. Since the thicknesses of TOx layers on both sides need to be precisely controlled, we developed a strategy to achieve the required precision which is described in supplementary information.

As a result, the RMS height of the silicon wafer was improved from 1.35 μm to 0.064 μm indicating an improvement factor of 21. The RMS slope error in the X direction was improved from 28 arcsecond to 0.862 arcsecond (32X) and in the Y direction from 16.9 arcsecond to 0.775 arcsecond (22X). The measured Zernike coefficients before and after correction are shown in Extended Data Table 2.

The residual error is dominated by an astigmatism term ($Z_{2-2}$), which could be due to the inaccurate FE modeling of the orthotropic substrate. A calibration process for verifying the data could significantly improve the precision of the corrections.

**Shape generation of trefoil and the secondary correction by Type-II mesostructured (Fig. 4a-Fig. 4d)**

The fabrication process for the Type-II mesostructure is similar to that for the calibration process. The following presents details of the pattern design.

$$\begin{bmatrix} \sigma_x \\ \sigma_y \\ \tau_{xy} \end{bmatrix} = \sum_{j=1}^{3} \begin{bmatrix} \cos^2(\theta_j) & \sin^2(\theta_j) & 2\cos(\theta_j)\sin(\theta_j) \\ \sin^2(\theta_j) & \cos^2(\theta_j) & -2\cos(\theta_j)\sin(\theta_j) \\ -\cos(\theta_j)\sin(\theta_j) & \cos(\theta_j)\sin(\theta_j) & \cos^2(\theta_j)-\sin^2(\theta_j) \end{bmatrix} \begin{bmatrix} \sigma_{uni,j} \\ 0 \\ 0 \end{bmatrix} \quad (8)$$

The equation above shows the principal stresses in a unit cell for the type-II mesostructures. The $\sigma_{uni,1}$ represents the uniaxial stress with the orientation of $\theta_1$ which is 0°. $\sigma_{uni,2}$ is the stress along 60°, and $\sigma_{uni,3}$ is along -60°. Based on the setting of the parameters, Eq. 8 can be written as

$$\begin{bmatrix} \sigma_x \\ \sigma_y \\ \tau_{xy} \end{bmatrix} = \begin{bmatrix} \sigma_{uni,1} + 1/4 \cdot (\sigma_{uni,2} + \sigma_{uni,3}) \\ 3/4 \cdot (\sigma_{uni,2} + \sigma_{uni,3}) \\ \sqrt{3}/4 \cdot (\sigma_{uni,3} - \sigma_{uni,2}) \end{bmatrix} \quad (9)$$

The stress field of trefoil can be expressed by

$$\begin{bmatrix} \sigma_x \\ \sigma_y \\ \tau_{xy} \end{bmatrix} = \begin{bmatrix} \cos^2(\phi) & \sin^2(\phi) & 2\cos(\phi)\sin(\phi) \\ \sin^2(\phi) & \cos^2(\phi) & -2\cos(\phi)\sin(\phi) \\ -\cos(\phi)\sin(\phi) & \cos(\phi)\sin(\phi) & \cos^2(\phi)-\sin^2(\phi) \end{bmatrix} \begin{bmatrix} \sigma \\ -\sigma \\ 0 \end{bmatrix} \quad (10)$$

$$\sigma = S_0 \times \frac{r}{R}, \quad \phi = \frac{1}{2}(\pi - \theta)$$

where $\sigma$ is the magnitude of the antibiaxial stress. The parameters $S_0$, $r$, $R$, $\phi$ and $\theta$ all have the same definition as in Eq. 3. By combining Eqs. 9 and 10, the uniaxial stresses required for generating the trefoil deformation are expressed as follows.

$$\begin{bmatrix} \sigma_{uni,1} \\ \sigma_{uni,2} \\ \sigma_{uni,3} \end{bmatrix} = \begin{bmatrix} 4/3 \cdot S_0 \cdot r/R \cdot (\cos^2(0.5 \cdot (\pi-\theta)) - \sin^2(0.5 \cdot (\pi-\theta))) \\ 2/3 \cdot S_0 \cdot r/R \cdot (\sin^2(0.5 \cdot (\pi-\theta)) - \cos^2(0.5 \cdot (\pi-\theta)) + 2\sqrt{3}\cos(0.5 \cdot (\pi-\theta))\sin(0.5 \cdot (\pi-\theta))) \\ 2/3 \cdot S_0 \cdot r/R \cdot (\sin^2(0.5 \cdot (\pi-\theta)) - \cos^2(0.5 \cdot (\pi-\theta)) - 2\sqrt{3}\cos(0.5 \cdot (\pi-\theta))\sin(0.5 \cdot (\pi-\theta))) \end{bmatrix} \quad (11)$$

In Eq. 11, the uniaxial stresses along the three directions need to be offset by a constant value to ensure compressive stress through the surface. The three offset stresses can be converted directly to the diameters of the TOx disks, as represented by C, D and E in Fig. 2e. The parameter $S_0$ is optimized so the diameter of each TOx disk does not exceed 500 μm which is the boundary of the grating disks.

After the patterning process (Fig. 4a), the measured deformation (Fig. 4b) shows that one side lobe of the trefoil is slightly higher than the others, which is created by an unexpected astigmatism term ($Z_{22}$) depicted in Fig. 4g. Since the $Z_{22}$ can be corrected by a constant uniaxial stress in the x direction (Eq. 2), we decided to perform a secondary adjustment by patterning voids at the center of selected TOx disks which have the grating lines in the X direction. Since the stress to be removed is simple, we followed a strategy of secondary correction without calculation, described in supplementary materials.

Since the stress reduction on the backside may lead to an increasing of spherical term ($Z_{20}$), additional steps may be followed to reduce the thickness of the TOx on the front side. However, the reduction of equibiaxial stress is not the major goal of this demonstration and these steps were thus omitted.

We eventually performed four iterations of secondary correction. The void diameters were increased from 50 μm to 115 μm to achieve the results in Fig. **4d**. The Zernike coefficients for each iteration are listed in the Extended Data Table 3. The target Zernike coefficient ($Z_{22}$) was successfully reduced from 44.58 nm to 2.97 nm, which successfully demonstrates the capability of secondary corrections on the Type-II mesostructure.

**Shape generation of the trefoil by Type-III mesostructured (Figs. 4e and 4f)**

The design rule of the Type-III mesostructure is similar to the one of the Type-II. However, the fabrication process is different. The process is described in supplementary materials.

# Extended Data

**Fig. 1. a Different Zernike polynomial zones require different types of stress to achieve deformation (a is reprinted from ref. 12)**

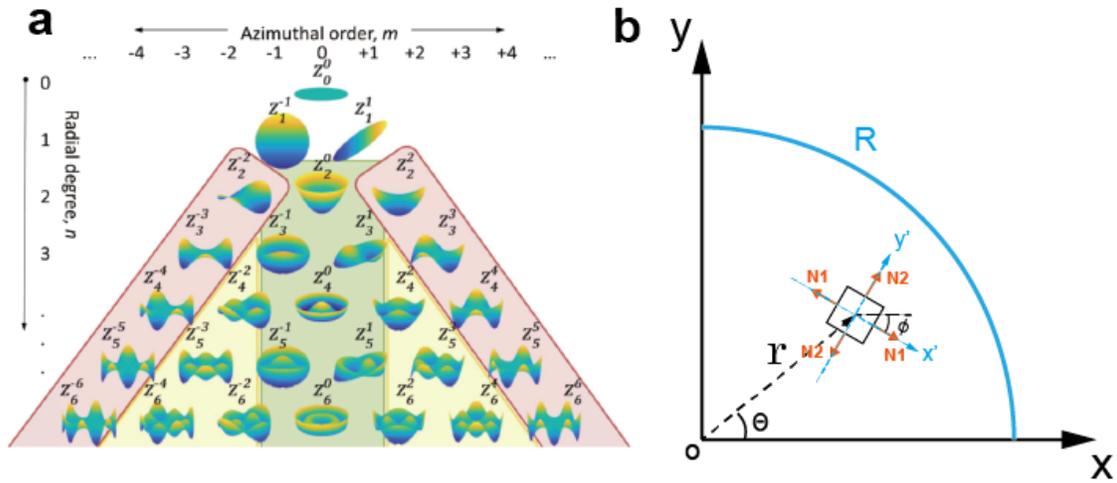

**a**, The center three columns of deformation (green zone) require equibiaxial stress fields. The pyramid shoulder (red zone) requires antibiaxial stress fields. The intermediate zone (yellow zone) requires a combination of both. **b**, $N_1$ and $N_2$ are the principal stresses with orientation of $\phi$. x-y and x'-y' indicate global and local Cartesian coordinate systems, respectively. In this work, R is the radius of the silicon wafer.

**Fig. 2. 2D FE modeling of deformations created by trenched lines on coated surfaces.**

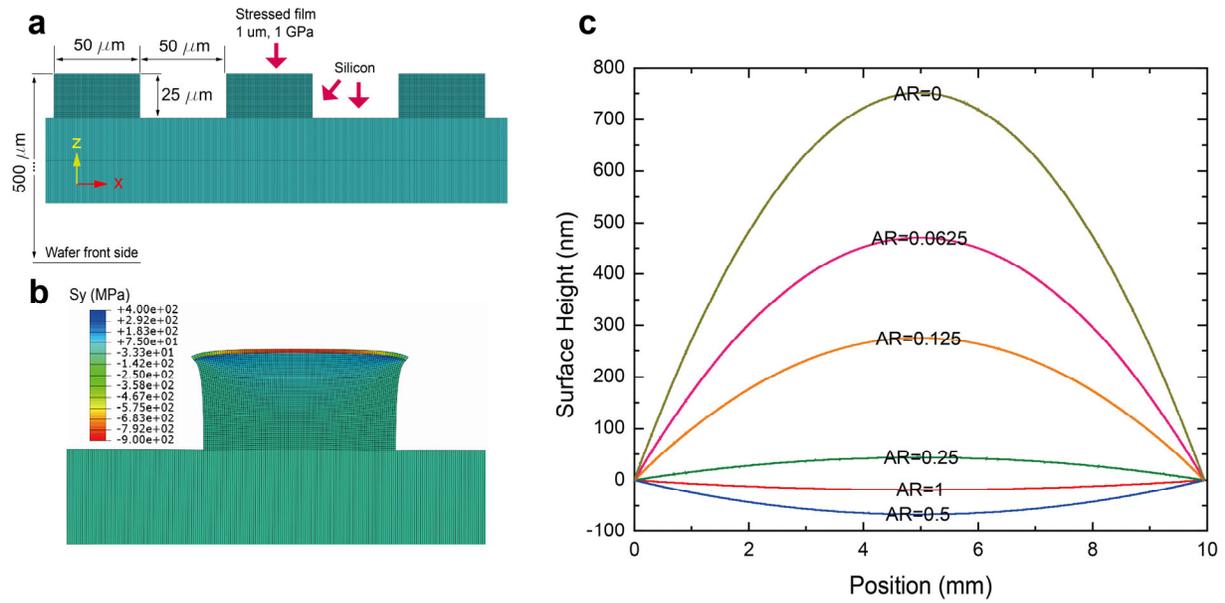

**a**, Sketch of a 2D model showing a cross section of trenched surface with AR=0.5. **b**, Calculated deformation and stress distribution of one tooth of the grating. Note that stress in the y direction is relaxed at the corners. **c**, Calculated deformation of wafer front sides for different ARs.

Note: Since the thickness of the 2D model is 0.525 mm which is way less than the length in the y direction (10 mm), the calculated curvature for different ARs (which is the key for deriving the red solid line in Fig. 1**b**) is not effected by the different size in the y direction between the model and real wafers.

**Fig. 3. Measured deformations generated by grating lines trenched into silicon with different aspect ratios.**

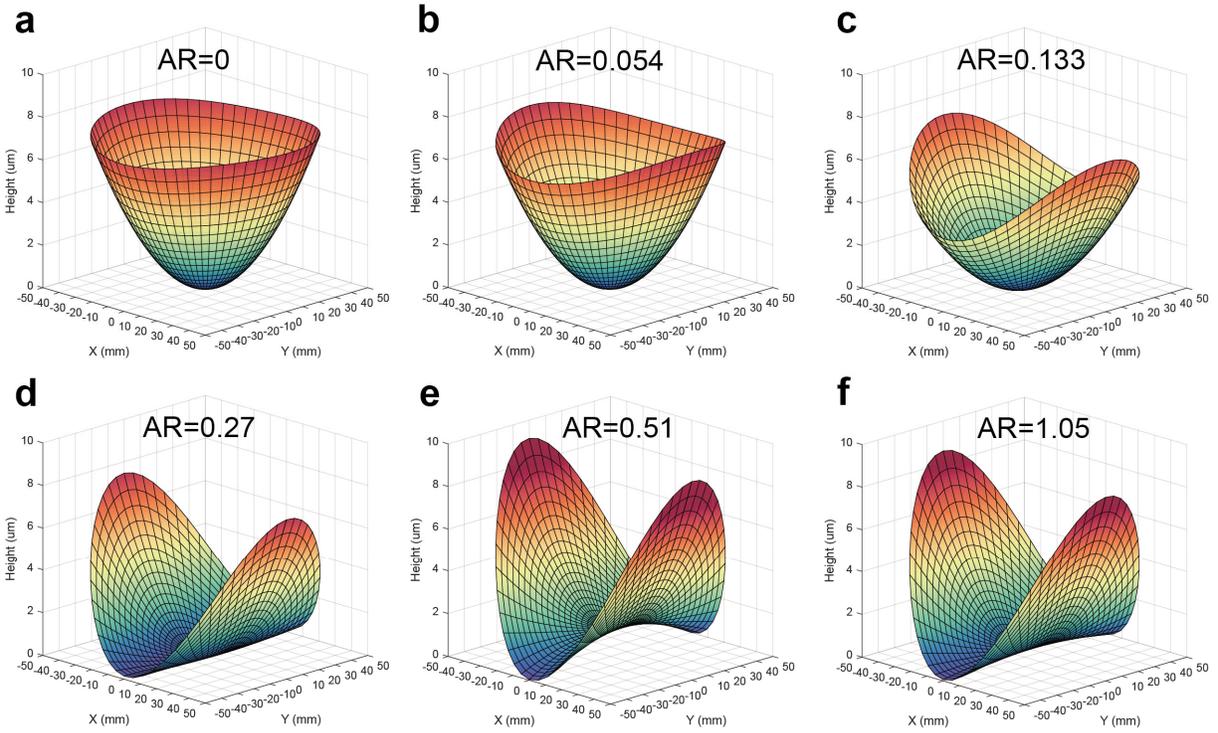

The pitch and the width of the lines are 10 μm and 3 μm, respectively. TOx on the back surface is 200 nm thick with no TOx on the front side. **a**, Trenches are only in TOx layer, AR=0. **b – f**, AR varies from 0.05 to 1.05.

**Fig. 4. Calibration of Type-I mesostructured patterns.**

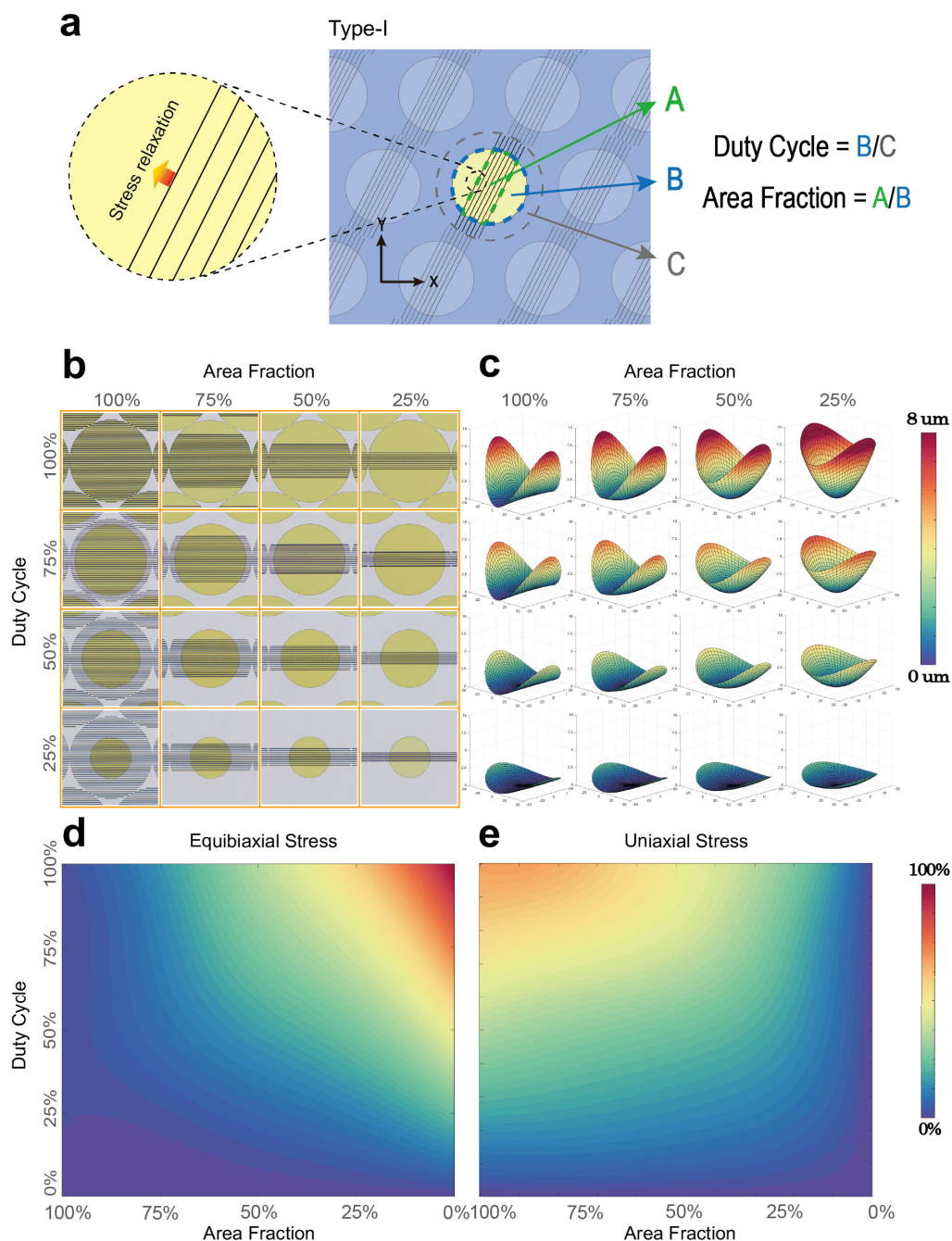

**a**, Right: Region A represents the grating trenched area encircled by dashed green line. Region B represents the TOx disk area enclosed by a dashed blue line. Region C indicates the unit circle area encircled by a dashed gray line. Left: Although the unpatterned TOx area is supposed to generate equibiaxial stress, the boundary of the trenched grating line can lead to a stress relaxation which influences the controlling precision. **b**, Microscope images of the unit cell regions on the backsides of 16 wafers. Each circle has different DC and AF values. **c**, Measured deformations of silicon wafers generated by the patterns in (**a**). **d** and **e**, Calibration maps of equibiaxial and uniaxial stresses for the two-dimensional variation of DC and AF. Note the maps were interpolated from 4 by 5 data points to 1000 by 1000 points.

**Fig. 5. Proposed Type-II mesostructure with two different stress providers**

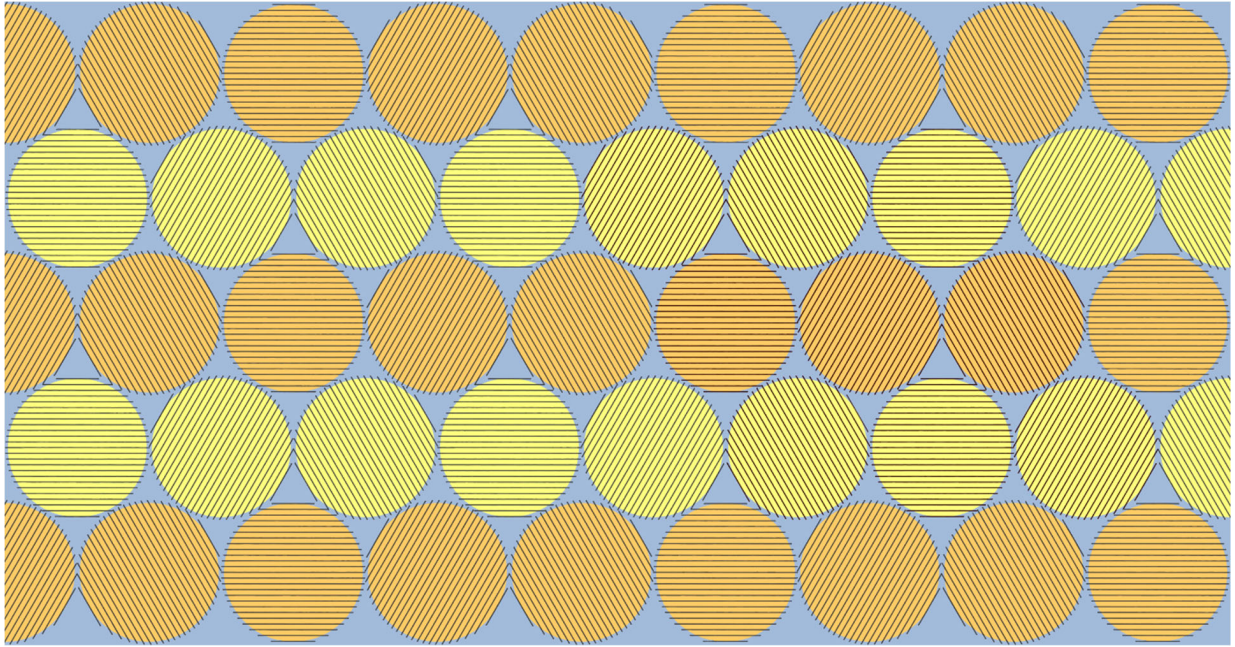

Orange regions represent coatings with tensile stress. Yellow represents compressive coatings.

**Fig. 6. Measured deformations of a patterned wafer when dipped in BOE for different times.**

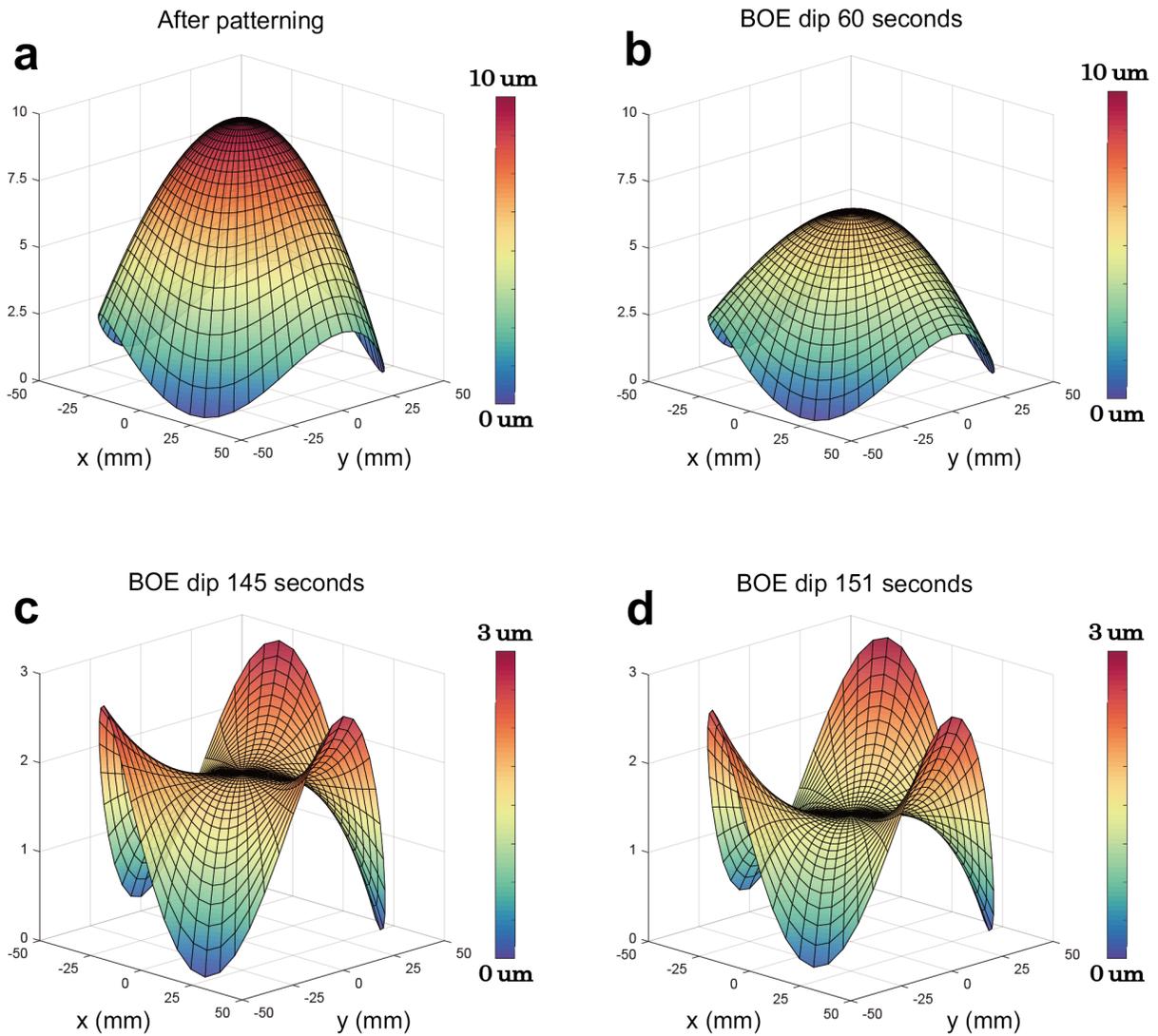

In the process, the patterned side is protected by PR. **a**, Deformation just after patterning the backside surface. **b**, Deformation when dipped in BOE for 60 seconds. **c**, After 145 seconds in total time, i.e., 85 seconds after (**b**). **d**, 151 seconds total time.

**Fig. 7. Calculated tensor fields for flattening a silicon wafer.**

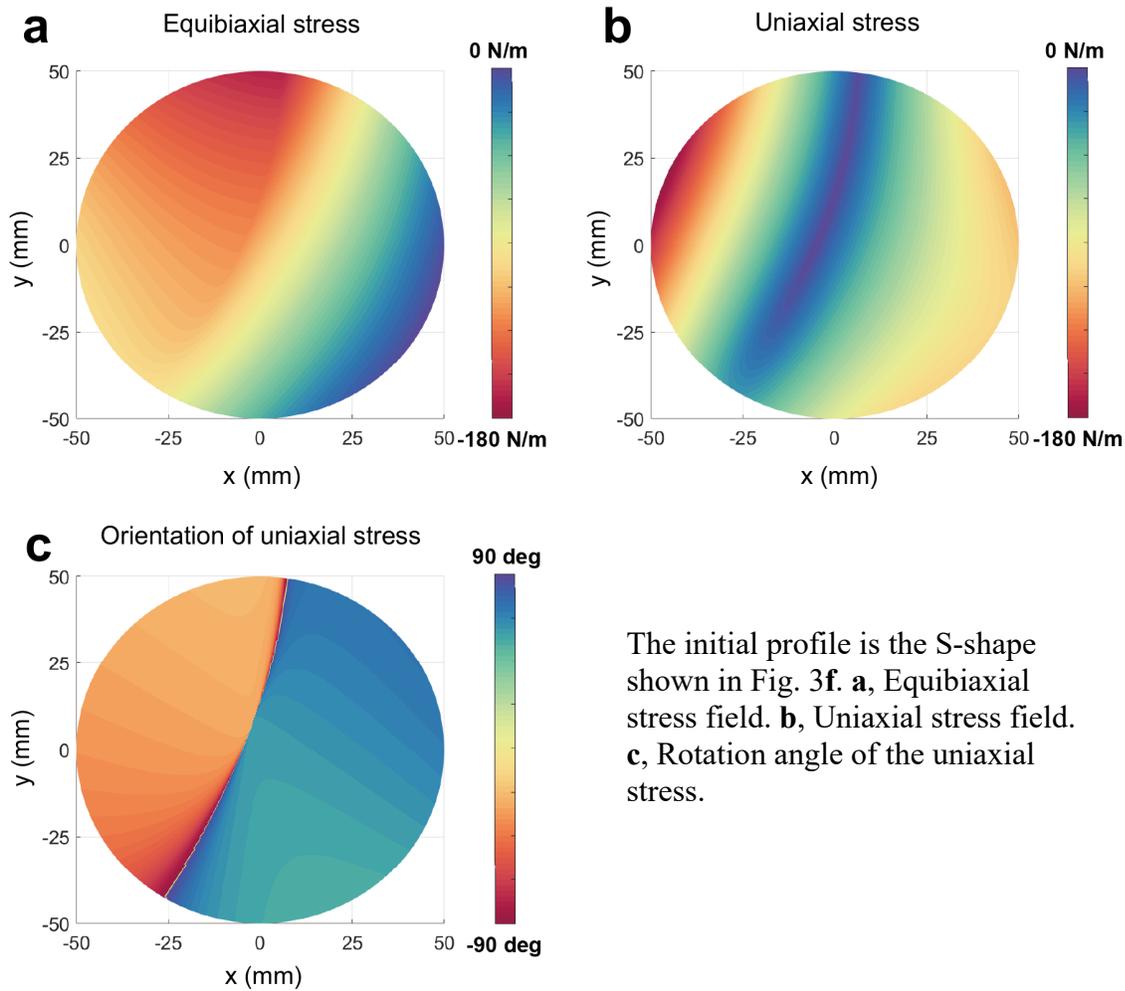

The initial profile is the S-shape shown in Fig. 3**f**. **a**, Equibiaxial stress field. **b**, Uniaxial stress field. **c**, Rotation angle of the uniaxial stress.

**Fig. 8. Definition of Ne and Na in Table S1.**

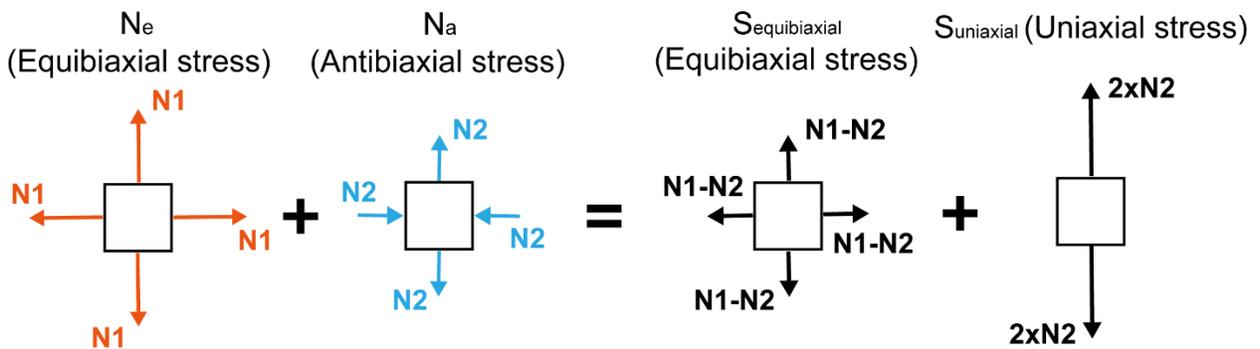

**Table 1. Stress fields for generating the deformation of each Zernike term with 100 nm RMS magnitude, derived from the analytical solution for isotropic substrates in ref. 12.** Ne, Na and Ns indicate equibiaxial stress, antibiaxial stress and shear stress represented by three Zernike polynomials. The explanation of $N_e$ and $N_a$ is shown in Extended Data Fig. 8.

| | | | $Z_{20}$ (100 nm) | | | $Z_{2-2}$ (100 nm) | | |
|---|---|---|---|---|---|---|---|---|
| Noll | n | m | Ne (N/m) | Na (N/m) | Ns (N/m) | Ne (N/m) | Na (N/m) | Ns (N/m) |
| 1 | 0 | 0 | -2.325 | 0 | 0 | 0 | 0 | -1.446 |
| 2 | 1 | 1 | 0 | 0 | 0 | 0 | 0 | 0 |
| 3 | 1 | -1 | 0 | 0 | 0 | 0 | 0 | 0 |
| 4 | 2 | 0 | 0 | 0 | 0 | 0 | 0 | 0 |
| 5 | 2 | -2 | 0 | 0 | 0 | 0 | 0 | 0 |
| 6 | 2 | 2 | 0 | 0 | 0 | 0 | 0 | 0 |

| | | | $Z_{22}$ (100 nm) | | | $Z_{3-1}$ (100 nm) | | |
|---|---|---|---|---|---|---|---|---|
| Noll | n | m | Ne (N/m) | Na (N/m) | Ns (N/m) | Ne (N/m) | Na (N/m) | Ns (N/m) |
| 1 | 0 | 0 | 0 | -1.446 | 0 | 0 | 0 | 0 |
| 2 | 1 | 1 | 0 | 0 | 0 | 0 | 0 | -2.505 |
| 3 | 1 | -1 | 0 | 0 | 0 | -5.695 | 2.505 | 0 |
| 4 | 2 | 0 | 0 | 0 | 0 | 0 | 0 | 0 |
| 5 | 2 | -2 | 0 | 0 | 0 | 0 | 0 | 0 |
| 6 | 2 | 2 | 0 | 0 | 0 | 0 | 0 | 0 |

| | | | $Z_{31}$ (100 nm) | | | $Z_{3-3}$ (100 nm) | | |
|---|---|---|---|---|---|---|---|---|
| Noll | n | m | Ne (N/m) | Na (N/m) | Ns (N/m) | Ne (N/m) | Na (N/m) | Ns (N/m) |
| 1 | 0 | 0 | 0 | 0 | 0 | 0 | 0 | 0 |
| 2 | 1 | 1 | -5.695 | -2.505 | 0 | 0 | 0 | -2.505 |
| 3 | 1 | -1 | 0 | 0 | -2.505 | 0 | -2.505 | 0 |
| 4 | 2 | 0 | 0 | 0 | 0 | 0 | 0 | 0 |
| 5 | 2 | -2 | 0 | 0 | 0 | 0 | 0 | 0 |
| 6 | 2 | 2 | 0 | 0 | 0 | 0 | 0 | 0 |

| | | | $Z_{33}$ (100 nm) | | | $Z_{40}$ (100 nm) | | |
|---|---|---|---|---|---|---|---|---|
| Noll | n | m | Ne (N/m) | Na (N/m) | Ns (N/m) | Ne (N/m) | Na (N/m) | Ns (N/m) |
| 1 | 0 | 0 | 0 | 0 | 0 | -9.004 | 0 | 0 |
| 2 | 1 | 1 | 0 | -2.505 | 0 | 0 | 0 | 0 |
| 3 | 1 | -1 | 0 | 0 | 2.505 | 0 | 0 | 0 |
| 4 | 2 | 0 | 0 | 0 | 0 | -10.397 | 0 | 0 |
| 5 | 2 | -2 | 0 | 0 | 0 | 0 | 0 | -6.468 |
| 6 | 2 | 2 | 0 | 0 | 0 | 0 | -6.468 | 0 |

| | | | $Z_{42}$ (100 nm) | | | $Z_{4-2}$ (100 nm) | | |
|---|---|---|---|---|---|---|---|---|
| Noll | n | m | Ne (N/m) | Na (N/m) | Ns (N/m) | Ne (N/m) | Na (N/m) | Ns (N/m) |
| 1 | 0 | 0 | 0 | -5.601 | 0 | 0 | 0 | -5.601 |
| 2 | 1 | 1 | 0 | 0 | 0 | 0 | 0 | 0 |
| 3 | 1 | -1 | 0 | 0 | 0 | 0 | 0 | 0 |
| 4 | 2 | 0 | 0 | -6.468 | 0 | 0 | 0 | -6.468 |

| 5 | 2 | -2 | 0 | 0 | 0 | -10.397 | 0 | 0 |
| 6 | 2 | 2 | -10.397 | 0 | 0 | 0 | 0 | 0 |
| | | | $Z_{44}$ (100 nm) | | | $Z_{4-4}$ (100 nm) | | |
| Noll | n | m | Ne (N/m) | Na (N/m) | Ns (N/m) | Ne (N/m) | Na (N/m) | Ns (N/m) |
| 1 | 0 | 0 | 0 | 0 | 0 | 0 | 0 | 0 |
| 2 | 1 | 1 | 0 | 0 | 0 | 0 | 0 | 0 |
| 3 | 1 | -1 | 0 | 0 | 0 | 0 | 0 | 0 |
| 4 | 2 | 0 | 0 | 0 | 0 | 0 | 0 | 0 |
| 5 | 2 | -2 | 0 | 0 | 4.573 | 0 | -4.573 | 0 |
| 6 | 2 | 2 | 0 | -4.573 | 0 | 0 | 0 | -4.573 |

**Table 2. Measured Zernike coefficients before and after correction (nm).**

| | $Z_{20}$ | $Z_{2-2}$ | $Z_{22}$ | $Z_{3-1}$ | $Z_{31}$ | $Z_{3-3}$ | $Z_{33}$ | $Z_{40}$ | $Z_{42}$ | $Z_{4-2}$ | $Z_{44}$ | $Z_{4-4}$ |
|---|---|---|---|---|---|---|---|---|---|---|---|---|
| Before | 184.06 | 334.09 | 287.18 | -537.99 | -606.09 | -958.03 | 9.39 | -14.84 | 14.13 | -45.85 | 52.6 | -199.1 |
| after | 12.4 | 56.95 | 7.74 | 0.94 | 11.1 | 10.17 | -9.42 | -4.26 | 3.19 | 5.23 | 14.02 | 10.15 |

**Table 3. Measured Zernike coefficients for the iterations of secondary corrections (nm).**

| No. | $Z_{20}$ | $Z_{2-2}$ | $Z_{22}$ | $Z_{3-1}$ | $Z_{31}$ | $Z_{3-3}$ | $Z_{33}$ | $Z_{40}$ | $Z_{42}$ | $Z_{4-2}$ | $Z_{44}$ | $Z_{4-4}$ |
|---|---|---|---|---|---|---|---|---|---|---|---|---|
| 1 | 97.19 | 9.46 | 44.58 | 2.36 | 16.14 | 12.82 | 121.33 | -7.71 | 0.55 | -6.92 | -1.69 | 5.41 |
| 2 | 107.33 | 11.58 | 34.30 | 1.90 | 11.76 | 16.43 | 127.30 | -12.80 | -2.12 | -1.98 | 2.32 | 1.84 |
| 3 | 88.97 | 5.33 | 15.26 | 12.48 | 17.52 | 10.66 | 125.33 | -11.32 | 2.47 | -3.78 | -2.65 | 6.22 |
| 4 | 93.92 | 13.13 | 2.97 | 5.22 | 13.92 | 15.06 | 125.09 | -12.56 | -1.38 | -6.88 | 1.05 | 6.26 |

**Acknowledgments:** We thank Prof. Jian Cao, Prof. Zheshen Zhang and Dr. William Zhang for helpful discussions.

**Funding:**

NASA Grant NNX14AE76G (MLS)

NASA Grant NNX17AE47G (MLS)

**Author contributions:**

Conceptualization: Youwei Yao

Experiment design, fabrication and metrology: Youwei Yao

Stress calculation: Brandon Chalifoux, Youwei Yao

Funding acquisition: Mark Schattenburg

Project administration: Mark Schattenburg

Supervision: Mark Schattenburg, Ralf Heilmann

Writing – original draft: Youwei Yao

Writing – review & editing: Brandon Chalifoux, Ralf Heilmann, Mark Schattenburg

**Competing interests:** Authors declare that they have no competing interests.

**Data and materials availability:** All data are available in the main text, the Extended Data and the supplementary informations.

## Supplementary informations

### Fabrication Process of the wafers in Extended Data Fig. 3.

1. Six new silicon wafers were selected and their front side topography measured as initial profiles.

2. Wafers were piranha cleaned, followed by a dry oxidation process (1060 °C, 4 hrs) to grow ~200 nm TOx on both sides as stressed coatings (approx. -70 N/m integrated stress).

3. The TOx layers were stripped from the front sides by using a buffered oxide etch (BOE). In this step, DOW SPR-700 photo resist (PR) was spin coated on the back sides to protect the stressed surfaces, and then removed by piranha after BOE.

4. Wafer front sides were measured with the TOx on the backside. By subtracting the initial profiles from Step 1, a spherical deformation is obtained, enabling calculation of the compressive stress in the TOx layers on each wafer based on the 3D FE model (38). In this step we assume the stresses in TOx layers are uniform.

5. 1 μm-thick PR (DOW SPR-700) layers were coated and baked on the wafer backsides for patterning purposes.

6. The PR-coated wafers were exposed by an MLA-150 patterning tool and developed in a developer solution (Microposit MF CD-26). Uniform grating lines with fixed pitch of 10 μm were created in the PR horizontally (x direction). The PR pattern duty cycle (the width of the PR tooth divided by the pitch of the grating lines which is 10 μm) is ~70% due to limitations of the exposure tool.

7. The wafers with patterned PR gratings were hard baked (110 °C, 1 hr) and then dipped in BOE for 3 min to remove the TOx within the unprotected areas. This isotropic wet etch does not impact the duty cycle since the TOx is much thinner than the width of the trench. The grating patterns have now been transferred into the 200 nm TOx layers.

8. The grating trenches were then etched into silicon by using a deep reactive ion etching (DRIE) tool. For each wafer the aspect ratio of the created grating teeth differs, varying from 0 to 1. This is achieved through control of the DRIE processing time. The first wafer has AR zero since it was not processed through DRIE.

9. The wafers were cleaned in piranha solution to remove residual PR.

10. The wafers with back-side grating lines were measured by the S-H metrology tool. Deformations are calculated and plotted (Extended Data Fig. 3) based on the initial profiles recorded in Step 1.

**The fabrication process for the 20 calibration wafers in Extended Data Fig. 4.**

1. 20 new silicon wafers were selected and their front sides measured as initial profiles.

2. Wafers were piranha cleaned, followed by a dry oxidation process (1060 °C, 4 hrs) to grow ~200 nm TOx on both sides as stressed coatings (approx. -70 N/m integrated stress).

3. The TOx layers were stripped from the front sides by using buffered oxide etch (BOE). In this step, SPR-700 photoresist (PR) was spin coated on the back sides and then removed by piranha after BOE to protect the stressed surfaces.

4. The topologies of the front sides were measured with the TOx on the backsides. By subtracting the initial profiles obtained from Step 1, spherical deformation is obtained enabling calculation of the compressive stress in the TOx layers on each wafer. In this step we assume the stress is uniform in the TOx layers.

5. 1 μm thick PR (DOW SPR-700) layers were spin coated and baked on the backsides of wafers for patterning purposes.

6. The PR-coated wafers were exposed with an MLA-150 patterning tool, developed in Microposit MF CD-26, etched by BOE and then piranha cleaned. The TOx disks with the different DCs patterned on the backsides of 20 wafers were then ready for further processing. Note that at this stage no grating lines have yet been trenched into the TOx.

7. A patterned wafer was chosen with DC=100%. The deformation was measured and the equibiaxial stresses field was calculated assuming the patterned TOx is a fictitious uniform film.

8. The calculated equibiaxial stress is compared to the stress fields derived in Step 4. The stress on each wafer for normalization is therefore derived.

9. The wafers were again spin coated with SPR-700, patterned by the MLA-150 and then developed in CD-26 to create grating patterns in the PR, accurately aligned with the TOx disks.

10. The wafers with PR patterns were hard baked (110 °C, 1 hr) and then dipped in BOE for 3 min to remove the TOx within the unprotected areas. The grating patterns have now been transferred to the 200 nm TOx layers.

11. The wafers with PR and TOx grating patterns are trenched into the silicon substrate using a deep reactive ion etching (DRIE) tool. For each wafer the aspect ratio of the created grating teeth is fixed (AR=1).

12. The wafers are piranha cleaned to remove residual PR.

13. The wafers now with Type-I mesostructured patterns are measured by the S-H metrology tool. Deformations are generated and plotted (Extended Data Fig. 4) based on the differences with the initial profiles recorded in Step 1. The equibiaxial stresses and uniaxial stresses are calculated and then normalized by the stresses derived in Step 8.

14. The normalized stresses are interpolated to 1000 by 1000 data points and plotted in Extended Data Figs. 4**d-e**.

**The iteration process for thinning the TOx film on the front side in Extended Data Fig. 6.**

1. The wafer's patterned side is spin coated by a thick layer of PR for protecting the trenched mesostructure.

2. The wafer is dipped in diluted BOE (1:3 - 1:10) so the TOx on the front side is etched uniformly and slowly. The etching time is controlled.

3. The wafer is piranha cleaned to strip the backside PR.

4. The deformation is measured. If the spherical term ($Z_{20}$) is not close to 0, recalculate the dipping time and repeat the process from Steps 1 to 4 until the profile is acceptable.

**The iteration process to to achieve the required precision on the samples in Fig. 3d-f.**

1. A wafer is selected and initial 2 μm-thick TOx layers are grown on both sides.

2. The wafer is dipped in BOE to reduce the thickness from 2 μm to 1.6 μm, slightly higher than the required thickness, so a stress margin is reserved for refined adjustments in the last steps.

3. The designed pattern is trenched into the backside. Since the thickness of the TOx is much larger than the ones we used for the calibration process and trefoil deformations, the patterning process is modified for better quality. A dry etcher (LAM-590) is used instead of BOE to trench the grating lines in the TOx layer.

4. After the patterning and cleaning processes, the surface profile of the deformed wafer is measured. At this stage, the spherical term ($Z_{20}$) should be under the correction target since the TOx on the front side is thicker than required. In the meantime, the remaining 11 Zernike terms ($Z_{2-2}$ to $Z_{4-4}$) are also over-corrected since the TOx on the backside is thicker than needed, providing higher compressive stress leading to over-deformation.

5. The front side is spin coated with PR. The wafer is dipped in diluted BOE for a controlled period so the thicknesses of the TOx layer on the backside is reduced.

6. The wafer is piranha cleaned and then measured by the metrology tool. The coefficients of the 11 Zernike terms should be close to zero.

7. If the flatness, excluding the spherical term, is not acceptable, then Steps 5 and 6 should be iteratively performed to achieve a good precision.

**The fabrication process of secondary correction on the Type-II samples in Fig. 4c**

1. Both sides of the patterned wafer were spin coated with PR (AZ5214).
2. The patterned surface was exposed by the MLA-150, developed, and BOE etched to create uniform voids in the TOx areas with grating lines in the X direction (Fig. 4**c**). In the first attempt, the void area was tiny for testing the secondary deformation (50 μm void's diameter).
3. The wafer is piranha cleaned and then measured for inspecting the Zernike coefficients.
4. If the suppression on $Z_{22}$ is not satisfied, repeat the Steps 1 to 3 with larger voids patterning until the profile is acceptable.

**The fabrication process on the Type-III samples in Fig. 4e**

1. A virgin silicon wafer is selected and front side topology measured as the initial profile.
2. PR (SPR-700) is spin coated on the front side.
3. The wafer is exposed by the MLA-150 and then developed by developer (CD-26) to create grating lines in the PR.
4. The wafer was trenched by deep reactive ion etch (DRIE) to transfer the patterns from PR into the silicon surface. The aspect ratio (AR) was set to 1 for testing purposes. The trenched silicon surface is depicted in the background of Fig. 4**e**.
5. The silicon wafers with grating line trenches are cleaned and thermally oxidized (1060 °C, 3 hrs) to grow ~200 nm TOx on the surfaces including the sidewall and adjacent areas of the grating teeth.
6. The oxidized wafer is then spin coated with PR (AZ-5214) on both sides.
7. The trenched side is exposed by the MLA-150 and developed in AZ-422 to create PR disk patterns overlapping the grating circles.
8. The wafer is dipped in BOE and followed by a piranha cleaning to transfer the PR pattern into the TOx. The patterned surface at this stage is shown in the foreground of Fig. 4**e**.